\documentstyle [12pt]{article}
\newcommand{\be}{\begin{equation}}
\newcommand{\ee}{\end{equation}}
\newcommand{\bees}{\begin{eqnarray}}
\newcommand{\ees}{\end{eqnarray}}
\newcommand{\ra}{\rightarrow}
\newcommand{\gsim}{\stackrel{>}{\sim}}
\newcommand{\lsim}{\stackrel{<}{\sim}}
\newcommand{\mpl}{M_{\rm Pl}}
\newcommand{\lpl}{L_{\rm Pl}}

\begin{document}

\begin{titlepage}
\begin{flushright}
 IFUP-TH 14/95\\
  April  1995\\
  gr-qc/9504020
\end{flushright}

\vspace{5mm}

\begin{center}

{\Large\bf  Effective Lagrangian for Quantum\\

\vspace{3mm}

Black Holes}

\vspace{12mm}
{\large Alessandra Buonanno$^{a,b}$, Mario Gattobigio$^{a}$,\\}

\vspace{2mm}

{\large Michele~Maggiore$^{b,a}$,   Luigi Pilo$^{c,b}$ } and
{\large Carlo Ungarelli$^{a,b}$}

\vspace{3mm}

(a) Dipartimento di Fisica dell'Universit\`{a},\\
piazza Torricelli 2, I-56100 Pisa, Italy.\\
(b) Istituto Nazionale di Fisica Nucleare, sezione di Pisa, Italy.\\
(c) Scuola Normale Superiore, Pisa, Italy.
\end{center}

\vspace{4mm}

\begin{quote}
\hspace*{5mm} {\bf Abstract.} We discuss the  most
general effective Lagrangian obtained from the assumption that the
degrees of freedom to be quantized, in a black hole, are on the
horizon. The effective Lagrangian depends only on the induced metric
and the extrinsic curvature of the (fluctuating) horizon, and the
possible operators can be arranged in an expansion in
powers of $\mpl/M$, where
$\mpl$ is the  Planck mass and $M$ the black hole mass.
We perform a  semiclassical expansion of the action with a formalism
which preserves general covariance explicitly. Quantum
fluctuations over the classical solutions are described by a
single scalar field living in the 2+1 dimensional world-volume swept by
the horizon, with a given coupling to the background geometry.
 We discuss the resulting field theory and we
compute the black hole entropy with our formalism.

\end{quote}
\end{titlepage}
\clearpage

\section{Introduction}
Effective Lagrangians are one of the most powerful tools of theoretical
physics. They allow us to investigate physics at large distances or
low energies when either the fundamental microscopic theory is unknown,
as is the case in quantum gravity, or when an explicit derivation of
large distance physics from the underlying fundamental theory is
technically very difficult, as, for instance, in {\em
QCD}. The only ingredients that are needed in an effective Lagrangian
approach are:  1) to know the symmetries of the underlying
theory, and to know how they are realized in the vacuum, and
2) to understand what are the relevant degrees of freedom
at low energies. These, in a non-perturbative regime,  are very
different from the fields which appear in the underlying Lagrangian,
and are in fact collective excitations of the fundamental variables.

An example which immediately comes to mind is the chiral Lagrangian
for pions.  In this case one knows
that the relevant degrees of freedom at low
energies are not the fundamental fields, quarks and gluons, but rather
the pion fields. Then, in order to write down the most
general effective Lagrangian, we do not need to  know that the
fundamental theory is {\em QCD}, but we only need to know how chiral
symmetry is realized. Where we pay for our  ignorance of short
distance physics is in the fact that we must introduce
phenomenological couplings, like $f_{\pi}$.
Then, if the microscopic theory is known, we can in principle derive
these couplings from it, although this is in general a difficult
non-perturbative problem.

The purpose of this paper is to set up a similar formalism for black
holes in quantum gravity.
The first question to be answered is: what are the relevant degrees of
freedom in terms of which we should write our effective Lagrangian? We
answer this question with the following

\begin{quote}
{\em Basic Postulate:} from the point of view of a static
observer, in a black hole the
degrees of freedom to be quantized are on the horizon. The
 effective Lagrangian is written in terms of the variables
$\zeta^{\mu}(\tau ,\sigma_1 ,\sigma_2)$ which define the position of
the quantum, fluctuating, horizon.
\end{quote}

The variables $\tau ,\sigma_1 ,\sigma_2$ parametrize the
world-volume swept by the horizon. For the Schwarzschild black hole,
we will usually fix the gauge $\tau =t,\sigma_1 =\theta ,\sigma_2
=\phi$, where $t$ is  Schwarzschild time and
$\theta ,\phi$ are the polar angles.

The symmetries to be respected by the effective Lagrangian are:
general covariance in  the embedding spacetime and
reparametrization invariance in the world-volume.

The above postulate, in our opinion, is in the spirit of ideas
of 't~Hooft~\cite{tH} and of Susskind and coworkers~\cite{Sus}.
In the framework of classical black holes, a
`membrane paradigm' was discussed in ref.~\cite{memb}.
Further investigations along similar lines have been recently presented
in refs.~[4-8].

In order to have a useful effective
Lagrangian we need, of course, a small parameter, so that it will be
possible to limit ourselves to the first few terms in the effective
theory.
Our small parameter, in the case of Schwarzschild black holes,
is $\mpl /M$, where $M$ is the black hole mass and $\mpl$ is the
Planck mass. We will show  that  higher dimension
operators in the effective Lagrangian  correspond to higher order
terms in an expansion in  powers of $\mpl /M$.

The paper is organized as follows.
In sect.~2 we show how to construct
 the most general effective Lagrangian compatible with
the basic postulate and with the symmetries
of the problem and we write down explicitly the lowest
dimension operators. In sect.~3,
using a generally covariant background field method,
we expand the action around the classical solutions and
we write it in terms of a single
scalar field $\phi (\tau ,\sigma_1,\sigma_2 )$
living in the world-volume. In sect.~4 we compute the black hole
entropy with this formalism, and we find corrections to the
relation $S=A/4$.
Sect.~5 contains the conclusions. Some technical issues are
discussed in Appendix A, while in
 Appendix B we  repeat our considerations
in the case of  the 2+1 dimensional black hole.

\clearpage

\section{The effective Lagrangian}

The fields $\zeta^{\mu}(\xi )$,  $\xi^i=(\tau ,\sigma_1 ,\sigma_2)$,
describe a 2+1 dimensional
timelike hypersurface (the world-volume)
embedded in 3+1 dimensional spacetime.
If $g_{\mu\nu}$ is
the   spacetime metric\footnote{We use the sign conventions of
Misner, Thorne and Wheeler~\cite{MTW}.}
 (to be specified below),
the intrinsic properties of the surface are
completely characterized by the induced metric $h_{ij}$,
\be
h_{ij}=g_{\mu\nu}\partial_i\zeta^{\mu}\partial_j\zeta^{\nu}\, ,
\ee
where $\partial_i=\partial /\partial\xi^i$,
and world-volume
indices take values $i=0,1,2$. The extrinsic properties
of the surface, instead, are completely characterized by the extrinsic
curvature tensor $K_{ij}$, defined by
\be\label{K}
K_{ij}=n_{\mu ;\nu}\partial_i\zeta^{\mu}\partial_j\zeta^{\nu}\, .
\ee
Here $n_{\mu}$ is the outer normal of
the surface $x^{\mu} =\zeta^{\mu}(\xi)$,
and the semicolon denotes as usual the covariant derivative in
the embedding space. The equations defining the normal $n^{\mu}$ are
\be\label{normal}
g_{\mu\nu}n^{\mu}\partial_i\zeta^{\nu}=0\, ,\hspace{5mm} i=0,1,2
\ee
together with the normalization condition
$g_{\mu\nu}n^{\mu}n^{\nu}= 1$. The
trace of the extrinsic curvature is $K=h^{ij}K_{ij}=K^i_i$
(in the following, world-volume indices are  raised or lowered
with the induced metric.)
We recall  that $K_{ij}$ is a
symmetric tensor (see e.g. ref.~\cite{HE}).
Since $h_{ij}$
determines completely the intrinsic properties of the surface while
$K_{ij}$ determines completely how the surface is embedded in 3+1
dimensional space, the effective action can  depend
only on these quantities, and must respect general covariance in
the embedding
spacetime and reparametrization invariance in the
world-volume.

Under parity transformations in the
world-volume (e.g. $\tau\ra\tau ,\sigma_1\ra\sigma_1,
\sigma_2\ra -\sigma_2$) the outer normal remains unchanged, and
therefore $K$ is a real scalar under these transformations.
In some special case, there is a symmetry transformation under which
$K$ transform as a pseudoscalar. This happens, for instance,
if we consider a planar membrane located at $z=0$, in ${\bf R}^4$.
In this case the reflection $z\ra -z$ is a symmetry, and
 this symmetry forbids the presence of odd powers of $K$ in the
effective action. However, in the general case, these terms will be
present.

Let us group the  possible terms that we can write down on the basis
of their dimensions. For the purpose of power counting, it is
convenient to assign dimensions of length to each of the $\xi^i$. Then
$h_{ij}$ is dimensionless and $K_{ij}$ has dimensions of mass (if we
use units $\hbar =c=1$ and we keep $G$ explicit).
The most general effective action compatible with our symmetries
can be written as
\be
S_{\rm eff}=
-{\cal T}\int d^3\xi\sqrt{-h}\, \left[ 1+\sum_a c^{(1)}_aO^{(1)}_a
+\sum_a c^{(2)}_aO^{(2)}_a+\ldots\right]\, .
\ee
Here ${\cal T}$ is the membrane tension and has dimensions of
(mass)$^3$.
The first term in this expansion is just the standard
Dirac membrane action~\cite{Dir},
i.e., the generalization to membranes of the Nambu-Goto
action for strings  and of the action $S=-m\int ds$ for pointlike
particles.
The operators $O^{(1)}_a$, enumerated by the index
$a$,  are all possible
operators with dimensions of mass, the operators $O^{(2)}_a$ have
dimensions of (mass)$^2$, etc. Correspondingly, the phenomenological
parameters $c^{(n)}_a$ have dimensions of (mass)$^{-n}$.

These operators must  be constructed with $K_{ij}$,
 $h_{ij}$ and their derivatives.
Because of reparametrization invariance
the derivatives of $h_{ij}$ will only
enter through its
Riemann tensor $R_{ijkl}$, which in
the $2+1$ dimensional world-volume  is fixed by the
Ricci tensor $R_{ij}$, and  through
the covariant derivatives of the Riemann
tensor. In the following $R_{ijkl},R_{ij},R$ will
always refer to the Riemann tensor, Ricci tensor and scalar curvature
of the world-volume, while script letters ${\cal
R}_{\mu\nu\rho\sigma}, {\cal R}_{\mu\nu},{\cal R}$ refer to
the embedding  space.
Since $K_{ij}$ has dimensions of mass
while $R_{ij}$ has dimensions of (mass)$^2$,
at level $n=1$ the only possible operator is $O^{(1)}=K$.
At the level $n=2$ we
have three different operators,
\be
O^{(2)}_1 = R,\hspace{3mm} O^{(2)}_2=K^2,\hspace{3mm}
O^{(2)}_3=K_{ij}K^{ij}\, .
\ee
However in the vacuum, where ${\cal R}_{\mu\nu}=0$,
they are not independent because of the Gauss-Codazzi relation,
$R=K^2-K_{ij}K^{ij}$. At order $n=3$ we have
\be
O^{(3)}_1 = R_{ij}K^{ij},\hspace{3mm} O^{(3)}_2=K^3,\hspace{3mm}
O^{(3)}_3=K^i_jK^j_lK^l_i,\hspace{3mm}
O^{(3)}_4 =K_{ij}K^{ij}K
\ee
where we have eliminated terms related   by the Gauss-Codazzi
relation  and  total derivatives.
 At order $n=4$, even if we make use of the Gauss-Codazzi relation,
 we have many independent terms:
\begin{eqnarray}
&& K^4, \hspace{1.5mm} K^2R, \hspace{1.5mm}
R^2, \hspace{1.5mm} R_{ij}R^{ij}, \hspace{1.5mm}
R_{ij}K^i_lK^{lj},\hspace{1.5mm} R_{ij}K^{ij}K,
\hspace{1.5mm} K^j_iK^l_jK^m_lK^i_m,\hspace{1.5mm}
KK^j_iK^l_jK^i_l, \nonumber \\
&& D_iK_{jl}D^iK^{jl},
\hspace{1.5mm}
D_iK_{jl}D^jK^{il},\hspace{1.5mm}
D_iKD^iK,\hspace{1.5mm}
D^iK_{ij}D^jK,\hspace{1.5mm}
D^iK_{ij}D_lK^{jl}\, ,
\end{eqnarray}
where $D_i$ is the covariant derivative in the world-volume.
Limiting ourselves to  order $n=2$,
and considering for the moment a  metric $g_{\mu\nu}$ generic,
we therefore write
\begin{eqnarray}
S_{\rm eff} & = & -{\cal T}\int d^3\xi\sqrt{-h}\, \left[ 1+
C_0 K+
C_{\scriptscriptstyle I}R+
C_{\scriptscriptstyle II}K^2
\right. \nonumber\\
\label{L2} & &
\left. + C_{\scriptscriptstyle III}K_{ij}K^{ij}
+({\rm  operators\,\, with\,} n\ge 3)\right]\, .
\end{eqnarray}
This effective action has  already been found by Carter and
Gregory~\cite{CG} in the context of domain walls in flat space. In
this case one has the 'microscopic' theory, which is the  theory
of a scalar field in a double well potential, and one can compute
explicitly the effective action for a thin domain wall. The result
of the explicit calculation turns out to be of the form~(\ref{L2}), in
agreement with the general arguments presented above.\footnote{In the
case of ref.~\cite{CG},
however, $C_0$ vanishes because the membrane divides the
embedding spacetime (in this case ${\bf R}^4$) into two identical
part. Then the reflection about the membrane is a symmetry of the
problem. Under this transformation $K\ra -K$ and therefore odd terms
are forbidden. This is not anymore the case if we consider, for
instance, a domain wall located at $z=0$ in a finite volume,
$-L_1\leq z\leq L_2$ with $L_1\neq L_2$. In this case a repetition of
the computation of ref.~\cite{CG} shows that $C_0\neq 0$.}

Dimensionally, we have $C_0=$ (length),
$C_{\scriptscriptstyle I},
C_{\scriptscriptstyle II}, C_{\scriptscriptstyle III}
=({\rm length})^{2}$. As in any
effective theory,  the scale for these coefficients is fixed by the
intrinsic cutoff of the effective Lagrangian, i.e. by the length scale
over which we have performed a coarse graining. For domain walls, in
fact, $C_{\scriptscriptstyle I},
C_{\scriptscriptstyle II}, C_{\scriptscriptstyle III}$
 turn out to be proportional to the square of
the thickness of the wall~\cite{CG}. In our case, as we will discuss in
sect.~3, the  membrane is a coarse graining
over a region of thickness $O(\lpl^2/R_S)$
or at most  $O(\lpl )$, where  $\lpl$ is the Planck length
and $R_S=2M$ is the Schwarzschild radius.
If superstrings are the fundamental theory we should  consider
the  string length rather then $\lpl$.
The two scales differ by a numerical factor
which is not very relevant for us at the moment. Thus,
$C_0$ is expected to be $O(\lpl^2/R_S)$ or at most $O(\lpl)$,
$C_{\scriptscriptstyle I},C_{\scriptscriptstyle II},
C_{\scriptscriptstyle III}$ are expected to
be $O(\lpl^4/R_S^2)$ or at most $O(\lpl^{2})$, etc.

The   length scale characterizing $K_{ij}, R_{ij}$ is instead the
curvature radius of the embedding spacetime. Thus, for dimensional
reasons, in the Schwarzschild metric
\be
K_{ij} =O\left( \frac{1}{R_S}\right)\, ,\hspace{5mm}
R_{ij} =O\left( \frac{1}{R_S^2}\right)\, .
\ee
Therefore the
expansion in higher dimension operators in the effective Lagrangian
is an expansion in powers of $(\lpl /R_S)^n$, with $n=2$ or at least
$n=1$. This is nothing but
the thin wall approximation used for domain walls.

A final ingredient for setting up our formalism is the choice of the
background metric $g_{\mu\nu}$. In the  limit of infinite
black hole mass any
backreaction on the classical metric due to
 the motion of the membrane is clearly neglegible, and we can simply
use the Schwarzschild  metric (or the Rindler metric, depending
on the case in which we are interested.)
Note that, since the basic postulate makes explicit reference to a
static observer outside the horizon, the Schwarzschild metric must
necessarily be expressed in Schwarzschild coordinates.
More in general, the metric
$g_{\mu\nu}$ should also include the backreaction of the membrane,
 and this will be a source of finite mass
corrections.

Note that on the Schwarzschild
and on the Rindler metric the three operators
appearing at  order $n=2$ are not independent, because of the
Gauss-Codacci relation, and therefore
we can limit ourselves to two of them.

\section{The semiclassical expansion}

The structure of the effective action can be greatly clarified by
expanding it around the classical solution of the equations of
motion. In this section and in sect.~4 we limit ourselves to the
leading term, i.e. to the Dirac membrane action.

The equations of motion
obtained by variation of the Dirac membrane action
can be written in terms of the trace of the extrinsic curvature as $K=0$.
In Rindler space we use coordinates $x^{\mu}=(t,x,y,z)$ and we fix the
gauge $\tau =t,\sigma_1 =x,\sigma_2 =y$. From
now on we use Planck units, setting  $G=1$.
The metric is
$g_{\mu\nu}={\rm diag}(-g^2z^2,1,1,1)$.
The Rindler metric is the limit of the Schwarzschild metric if we
send $M\ra\infty$, while remaining at a fixed, limited, distance from
the horizon, if  $g$ is identified with $1/(4M)$.
If we look for planar solutions of
the form $z=z(\tau )$, the equation of motion $K=0$ takes
the form~\cite{MM1}
\be\label{eqR}
z\ddot{z}-2\dot{z}^2+g^2z^2=0\, ,
\ee
and has the solution
\be\label{sol}
z_{\rm cl}(\tau )=\frac{z_0}{\cosh g\tau}\, .
\ee
For Schwarzschild black holes, we use coordinates $x^{\mu}=(t,r,\theta
,\phi)$ and $g_{\mu\nu}={\rm diag}(-\alpha
,\alpha^{-1},r^2,r^2\sin^2\theta )$, with $\alpha =1-(2M)/r$.
In order to facilitate the comparison with the Rindler limit, we use
the notation $g=1/(4M)$ and we define $z=\alpha^{1/2}/g$. The equation
of motion of a spherical membrane,
 written in terms of $z(\tau )$, is
\be\label{eqS}
z\ddot{z}-2\dot{z}^2+g^2z^2(1+3g^2z^2)(1-g^2z^2)^3=0\, .
\ee
This equation can be integrated exactly in terms of the
elliptic function of
the third kind and the explicit expression is quite complicated. However,
one can observe that, if $|\tau |\ra\infty$, the solution approaches
the horizon and therefore $gz(\tau )$ is small. Then the
equation can be solved perturbatively; the non linear terms
in eq.~(\ref{eqS}) can be neglected and the solution reduces to
 eq.~(\ref{sol}).

Now we can expand the action considering
fluctuations around a classical solution of the
equations of motion. The expansion can be performed in a way which
preserves  general covariance explicitly. The technique is basically the
same which was used in the classical works on the background field
method for the nonlinear $\sigma$-model~\cite{sigma}, and in the case
of domain walls it has been discussed in refs.~\cite{GV,Guv}.

One starts with the observation~\cite{GV}
that fluctuations along the surface
are 'pure gauge' and can be
reabsorbed with a reparametrization. The only physical quantum
fluctuations
are the ones perpendicular to the surface, and can be parametrized by a
single scalar field $\phi (\xi )$ living in the world-volume.
Let us denote by
$\bar{n}^{\mu}(\xi )$  the normal to the classical solution
$\bar{\zeta}^{\mu}(\xi )$.
In a generic coordinate system, for any given value of $\xi$ we
consider the (spacelike) geodesic $\zeta^{\mu}_{\rm geod}(\phi ,
\bar{\zeta}, \bar{n})$ parametrized by an affine parameter
$\phi$, which at $\phi =0$ goes through the point
$\bar{\zeta}^{\mu}$, with a tangent at $\phi =0$ equal to
$\bar{n}^{\mu}$. Then one writes~\cite{sigma}
\be\label{curved}
\zeta^{\mu}(\xi )=\zeta^{\mu}_{\rm geod}\left(
\phi (\xi ),\bar{\zeta}(\xi ), \bar{n}(\xi )
\right)\, .
\ee
Expanding the geodesic equation in powers of $\phi$ one gets
\be\label{geo}
\zeta^{\mu}=\bar{\zeta}^{\mu}+\phi\bar{n}^{\mu}-\frac{\phi^2}{2}
\bar{\Gamma}_{\rho\sigma}^{\mu}\bar{n}^{\rho}\bar{n}^{\sigma}
-\frac{\phi^3}{3!}
(\overline{\partial_{\tau}
\Gamma}_{\rho\sigma}^{\mu}-
\bar{\Gamma}_{\tau\rho}^{\alpha}\bar{\Gamma}_{\alpha\sigma}^{\mu} -
\bar{\Gamma}_{\tau\sigma}^{\alpha}\bar{\Gamma}_{\alpha\rho}^{\mu})
\bar{n}^{\rho}\bar{n}^{\sigma}\bar{n}^{\tau}+
\ldots\, ,
\ee
where the overbar denotes the value at $\bar{\zeta}$.
In order to expand the action in powers of $\phi$ it is extremely
convenient to use Riemann normal coordinates. In
these coordinates the  geodesics are straight lines, which means that
in these coordinates
 $\bar{\Gamma}_{\rho\sigma}^{\mu}=0$ and all the derivatives
$\partial_{\tau_1}\ldots\partial_{\tau_n}
\Gamma_{\rho\sigma}^{\mu}$
vanish  when  evaluated at $\bar{\zeta}$ and
fully symmetrized in the  indices $(\tau_1,\ldots
,\tau_n,\rho ,\sigma )$.
Then eq.~(\ref{geo}) becomes
\be
\zeta^{\mu}(\xi )=\bar{\zeta}^{\mu}(\xi )+\phi(\xi )\bar{n}^{\mu}(\xi
)\, .
\ee
The expansion
for $\partial_i\zeta^{\mu}$ can be evaluated by taking the
first derivative of eq.~(\ref{geo}), and using the fact that in
Riemann normal coordinates
\be
\overline{\partial_{\nu}\Gamma}_{\rho\sigma}^{\mu} =
-\frac{1}{3}({\bar{{\cal R}}^{\mu}}_{\hspace*{1.5mm}\rho\sigma\nu}+
{\bar{{\cal
R}}^{\mu}}_{\hspace*{1.5mm}\sigma\hspace*{-0.2mm}\rho\nu})\, ,
\ee
with similar relations for higher order derivatives~\cite{Petrov}.
Then at quadratic order one has
\be\label{part}
\partial_i\zeta^{\mu}=\partial_i\bar{\zeta}^{\mu}+
\partial_i(\phi\bar{n}^{\mu})+\frac{\phi^2}{3}
{\bar{{\cal R}}^{\mu}}_{\hspace*{1.5mm}\rho\sigma\alpha}
\bar{n}^{\rho}\bar{n}^{\sigma}
\partial_i\bar{\zeta}^{\alpha}+O(\phi^3)\, .
\ee
At the same time, in Riemann normal coordinates the metric $g_{\mu\nu}$
has an expansion in terms of the Riemann tensor and its
covariant derivatives
which at quadratic order reads
(see e.g. ref.~\cite{Petrov}),
\be\label{gmunu}
g_{\mu\nu}(\zeta )=\bar{g}_{\mu\nu}-\frac{\phi^2}{3}
\bar{\cal{R}}_{\mu\rho\nu\sigma}\bar{n}^{\rho}
\bar{n}^{\sigma}+O(\phi^3)\, .
\ee
Using eqs.(\ref{part},\ref{gmunu})
we can compute the expansion of the induced metric
$h_{ij}=g_{\mu\nu}(\zeta )\partial_i\zeta^{\mu}\partial_j\zeta^{\nu}$ and
therefore of $\sqrt{-h}$. The term linear in $\phi$ is simply
$\phi \bar{K}$
and vanishes because of the equation of motion $\bar{K}=0$,
and at quadratic
order one gets (after reabsorbing a factor ${\cal T}^{-1/2}$ into the
definition of $\phi$)
\be\label{lin}
S_{\rm eff}=S_{\rm cl}-\frac{1}{2}\int d^3\xi \sqrt{-\bar{h}}\left[
\partial_i\phi\partial^i\phi - (\bar{K}_i^j\bar{K}_j^i+
\bar{{\cal R}}_{\mu\nu}\bar{n}^{\mu}\bar{n}^{\nu})\phi^2\right]\, ,
\ee
where $S_{\rm cl}=-{\cal T}\int d^3\xi (-\bar{h})^{1/2}$,
and the indices are raised and lowered with $\bar{h}_{ij}$.
Eq.~(\ref{lin}) agrees with the result found by Carter with more
geometrical methods~\cite{Carter}.

Eq.~(\ref{lin}) is  the  action of a scalar field in curved space
in 2+1 dimensions. For a planar membrane in Rindler space,
$\bar{\zeta}^{\mu}=(\tau,x,y,z_{\rm cl}(\tau ))$
with $z_{\rm cl}$ given in eq.~(\ref{sol}). Then
\be\label{ind}
\bar{h}_{ij}={\rm diag}(-g^2z_{\rm cl}^2+\dot{z}_{\rm cl}^2,1,1)=
{\rm diag}(-\frac{g^2z_0^2}{\cosh^4g\tau} ,1,1)\, .
\ee
and $\bar{K}_i^j\bar{K}_j^i=0,{\cal R}_{\mu\nu}=0$.
Thus eq.~(\ref{lin}) describes
 a  massless scalar field living in a 2+1 dimensional space with
\be\label{back}
ds^2=-\frac{g^2z_0^2}{\cosh^4g\tau}d\tau^2 +dx^2+dy^2\, .
\ee
Introducing $\tilde{\tau}=z_0\tanh g\tau$, it becomes a flat metric
with boundaries in the time direction, since
$-z_0<\tilde{\tau}<z_0$.
The propagator of the $\phi$ field in this background
is discussed in Appendix A.

Let us consider now the Schwarzschild background.
The membrane represents a coarse graining over the region of space
where the local Hawking temperature reaches the scale where unknown
physics sets in. This certainly happens at least at the Planck
mass, or at $T=O(1)$ in Planck units. From
the expression of the local temperature near the horizon,
\be
T=\frac{1}{8\pi M\sqrt{-g_{00}}}\simeq \frac{\sqrt{2 M}}{8\pi
M\sqrt{r-2M}}\, ,
\ee
we see that $T\gsim O(1)$
when $r-2M\lsim O(1/M)$. Thus, this is the region
over which we have to perform a coarse graining, if we do not
want to be faced with physics beyond the Planck mass.
This corresponds to considering membranes whose area is
 $A=4\pi R_S^2+\delta A$, with $\delta A\sim 4\pi\lpl^2$. Note that,
instead, when $r-2M$ is on the order of a few Planck lengths, the local
Hawking temperature is  $\sim M^{-1/2}$ and it is still
small for black holes with large $M$.

Then, let us study the theory obtained by expanding around a spherical
solution of the equations of motion with
initial conditions $r(0)=2M+\delta r$,
$\dot{r}(0)=0$. The minimum value of $\delta r$ in which we are
interested is $\delta r=O(1/M)$, since at this scale we enter the
super-Planckian region. The largest value of interest are instead
$\delta r=O(1)$, since this is the scale where quantum effects on the
horizon start to become important~\cite{MM}.
In terms of the variable $z=\alpha^{1/2}/g$, using the notation
$z_0=z(\tau =0)$, this gives
 $z_0$ between $O(1)$ and $O(M^{1/2})$ and
therefore $gz_0$ between $O(1/M)$ and $O(1/M^{1/2})$.
Since $gz(\tau )\leq gz_0\ll 1$, the equation of
motion for a spherical membrane, eq.~(\ref{eqS}), can be solved
perturbatively in $gz_0$ and to lowest order it
reduces to the Rindler equation of motion.
Then, to lowest order in $gz_0$, the interval in
the $2+1$ dimensional world-volume can be written as
\be\label{sch}
ds^2 \simeq -\frac{g^2z_0^2}{\cosh^4 g\tau}d\tau^2+R_S^2
(d\theta^2 +\sin^2\theta\, d\phi^2 )\, .
\ee
As $\tau\ra\infty$  any deviation of a classical solution from spherical
symmetry  is washed out exponentially~\cite{MM1},
so the result obtained expanding around a non-spherical
background are qualitatively similar.

The main difference with the Rindler case is that now the mass term in
the Lagrangian, eq.~(\ref{lin}), is non zero. Strictly speaking, the
quadratic term is not exactly a mass term, since it depends on
$\tau$. However, it has a finite limit for $\tau\ra\pm\infty$ (which
is independent of whether we take an exactly spherical configuration
or not) which, to lowest order in $\alpha_0=(gz_0)^2$, is
\be
\bar{K}_i^j\bar{K}_j^i\ra \frac{6\alpha_0}{R_S^2}
\ee
On the
Schwarzschild metric ${\cal R}_{\mu\nu}=0$ and therefore it does not
contribute to the mass term.
Thus,  similarly to what happens for domain walls
in flat space~\cite{GV}, the mass term has a tachyonic sign.
Let us consider the evolution of a generic perturbation
$\phi (\tau ,\theta ,\phi)$. We
expand the perturbation in spherical harmonics,
\be
\phi (\tau ,\theta ,\phi)=\sum_{lm}\phi_{l}(\tau )Y_{lm}(\theta ,\phi)
\ee
and we define
\be
m^2=\frac{6\alpha_0}{R_S^2}\, .
\ee
In the range of values of $r(0)$ which is relevant for us, $m^2$
varies between $O(1/M^4)$ and $O(1/M^3)$.
The equation of motion of the perturbation in the background  given by
eq.~(\ref{sch}) is
\be
\left[ \frac{\partial^2}{\partial\tilde{\tau}^2}-
m^2+\frac{l(l+1)}{R_S^2}\right]\phi_l (\tau)=0\, ,
\ee
with $\tilde{\tau}=z_0\tanh g\tau$.
In the large $M$ limit only the mode with $l=0$ is unstable, and
 solving the equation of motion for the fluctuations at quadratic
order, we get
\be
\phi_0 (\tau )=\phi_0 (0)e^{\pm m\tilde{\tau}}=
\phi_0 (0)\exp\{ \pm mz_0\tanh g\tau\}\, .
\ee
As $\tau\ra\infty$ the unstable mode grows, but only increases up to the
finite value $\phi_0 (0)\exp\{ mz_0\}$. In the range of values of
$z_0$ which is more interesting for us,
 $mz_0$ varies between $O( 1/M^2)$ and $O(1/M)$ and therefore the
fluctuations only increase up to a small value.

We have  computed higher order terms in the potential part of the
Lagrangian, i.e., in the part which does not involve derivatives of
$\phi$. Writing
\be
V(\phi )=-\frac{1}{2}m^2\phi^2 +v_3\phi^3+v_4\phi^4+\ldots \,
\ee
we find, for a generic background,
\be
v_3=\frac{1}{3}\bar{n}^{\rho}\bar{n}^{\sigma}
\bar{{\cal
R}}_{\mu\rho\nu\sigma} \bar{K}^{\mu\nu}+
\frac{1}{3}\bar{K}^i_j\bar{K}^j_l\bar{K}^l_i
-\frac{1}{6}\bar{n}^{\rho}\bar{n}^{\sigma}\bar{n}^{\tau}
{\bar{{\cal R}}}_{\rho\sigma ;\tau}
\, ,
\ee
\bees
v_4&=&\frac{1}{12}\bar{n}^{\rho}\bar{n}^{\sigma}\bar{n}^{\tau}
\bar{K}^{\nu}_{\mu}
{\bar{{\cal R}}^{\mu}}_{\hspace*{1.5mm}\rho\nu\sigma ;\tau}
-\frac{1}{12}\bar{n}^{\alpha}\bar{n}^{\beta}\bar{n}^{\gamma}
\bar{n}^{\delta}
\bar{{\cal R}}^{\rho}_{\hspace*{1.5mm}\delta\tau\alpha}
\bar{{\cal R}}^{\tau}_{\hspace*{1.5mm}\beta\rho\gamma}\nonumber\\
 & & -\frac{1}{3}\bar{n}^{\rho}\bar{n}^{\sigma}\bar{K}^{\alpha}_{\mu}
\bar{K}^{\nu}_{\alpha}\bar{{\cal R}}^{\mu}_{\hspace*{1.5mm}\rho\nu\sigma}
-\frac{1}{4}\bar{K}^i_j\bar{K}^j_l\bar{K}^l_m\bar{K}^m_i
+\frac{1}{8}(\bar{K}^i_j\bar{K}^j_i)^2\\
& & -\frac{1}{12}\bar{n}^{\mu}\bar{n}^{\nu}\bar{n}^{\rho}\bar{n}^{\sigma}
{\bar{{\cal R}}}_{\mu\nu ;\rho\sigma}+
\frac{1}{8}(\bar{n}^{\mu}\bar{n}^{\nu}{\bar{{\cal R}}}_{\mu\nu})^2
+\frac{1}{4}\bar{n}^{\mu}\bar{n}^{\nu}{\bar{{\cal R}}}_{\mu\nu}
\bar{K}^i_j\bar{K}^j_i
\, ,\nonumber
\ees
where $K_{\mu\nu}=(\delta_{\mu}^{\alpha}+n_{\mu}n^{\alpha})n_{\nu
;\alpha}$ is the extrinsic curvature in 4-dimensional notation.
Evaluating these quantities on the Schwarzschild metric,  taking the
limit $\tau\ra\pm\infty$ and limiting ourselves to the leading term in
$\alpha_0$ we get
\bees
v_3&\ra& -\frac{\alpha_0^{1/2}}{R_S^3}\\
v_4&\ra& -\frac{1}{8R_S^4}\, .
\ees
Thus, also the cubic and quartic term give a negative contribution to
the potential, which (without reabsorbing ${\cal T}$ into $\phi$)
can be written as
\bees
V(\phi )&=&-\alpha_0^2{\cal T}\left[ 3\varphi^2 +\varphi^3
+\frac{1}{8}\varphi^4+\ldots\right]\, ,\\
\varphi &{\stackrel{{\rm def}}{=}}& \frac{\phi}{\alpha_0^{1/2}R_S}\, .
\ees
The fact that the quadratic  term in the
potential have a negative sign shows that, at the quantum level, $\phi
=0$ is not the true ground state of the theory. The problem is due to
the spherical mode, $l=0$, since for all higher modes the term
$l(l+1)/R_S^2$ dominates over $m^2$, if $M$ is large.

The spherical mode, however, has been treated exactly, i.e. without
performing a semiclassical expansion, in ref.~\cite{MM1}, where it has
been obtained  the Schroedinger equation and therefore
the wave function of a spherical membrane. The resulting distribution
probability of the radial mode is peaked at a non-zero value of
$r-R_S$, which depends also on the membrane tension.

The fact that in the semiclassical  expansion the spherical mode is
unstable therefore reflects an intrinsic limitation of the semiclassical
approximation. Another, related, limitation is due to the fact that
the solution $\zeta^{\mu}_{\rm cl}$
of the classical equation of motion approaches
asymptotically the horizon. However, the variables $\zeta^{\mu}$
emerge only after performing a coarse graining over the appropriate
scale of distances, and therefore in their definition is implicit an
uncertainty on the order of this length scale. Therefore, it is not
legitimate to extrapolate the solution of the equation of motion down
to distances very close to the horizon. When the classical solution
$\zeta^{\mu}_{\rm cl}$
formally reaches a value of $r= R_S+O(\lpl^2/R_S)$, it
would be physically more sensible to use, instead of $r_{\rm
cl}(\tau )$, a
membrane essentially static at an average value of $r=
R_S+a\lpl^2/R_S$, with $a$ some numerical constant. This approach was
recently  discussed by Lousto and one of the authors~\cite{LM}.
In this case one gets again a Klein-Gordon equation for the
 fluctuations, in terms of a variable $\tilde{\tau}$ which
now is defined as $\alpha_M^{1/2}\tau$, where $\alpha_M$ is the value
of $\alpha$ at the average membrane location. Therefore $\tilde{\tau}$
is just the local time, and the energy conjugate to it is
 the local energy.

\section{Black hole entropy}

\subsection{The entropy of the $\phi$ field}
In our approach, the horizon is described by a scalar field living
in the horizon world-volume, and the microscopic degrees of freedom of
the black hole (or its 'quantum hair') are the modes of this
field. Their contribution to the entropy can be estimated as follows.

 Let us consider first the Rindler
metric. In this case the field $\phi$ is massless, and
therefore we have to compute the entropy of a massless boson gas  in
2+1 dimensions.
Let us call $A=L^2$ the area of the horizon. For
Rindler space this should be eventually   sent to infinity, so what
matters is the entropy per unit area.
The modes of the field $\phi$
are labeled by ${\bf k}=(k_x,k_y)$ and
the free energy $F$ of a
2+1 dimensional massless boson gas taken at the
inverse temperature $\beta$ is
\be
\beta F=\sum_{\bf k}\log (1-e^{-\beta E})\, ,
\ee
with $E=|{\bf k}|$. For large $L$ the summation over modes can be
replaced with an integration, using
\be
\sum_{\bf k}=\left(\frac{L}{2\pi}\right)^2\int d^2k\, .
\ee
The integral over $k$ converges  both at small and at large values of
$k$. However,
physically it does not make much sense  to include in the integral
over $d^2k$ modes of the $\phi$ field with arbitrarily high energy.
Then we get for the free energy and the entropy
\begin{eqnarray}
F&=&-\frac{c(\Lambda )}{2\pi\beta^3}\, A\\
S&=&-\left(\frac{\partial F}{\partial T}\right)_A
=\frac{3c(\Lambda )}{2\pi\beta^2}\, A\, ,\label{entr}
\end{eqnarray}
where
\be
c(\Lambda )=-\int_0^{\Lambda}dx\, x\log (1-e^{-x})\, ,
\ee
and $\Lambda =\beta E_{\rm max}$.
If $\Lambda\ra\infty$ then
 $c(\Lambda )\ra\zeta (3)\simeq 1.2$,
where $\zeta (x)$ is the Riemann zeta function. With a finite cutoff
we get a smaller value. For instance, $c(1)\simeq 0.4$.

The same result is obtained for a Schwarzschild black hole in the
large mass limit. In fact for Schwarzschild black holes in the
large $M$ limit we still have massless bosons,
 since the (tachyonic) mass term is $\sim M^{-2}$. The modes are
labelled by the angular momentum quantum numbers $(l,m)$, and
\be
\beta F=\sum_{l=0}^{\infty}(2l+1)\log\left(1-e^{-\beta E_l}\right)\, ,
\ee
\be
E_l=\frac{\sqrt{l(l+1)}}{R_S}\, .
\ee
For large $R_S$ the spacing between the levels is small and we can
approximate the sum over $l$ with an integral. After an integration by
parts we get
\be
 F=-\frac{1}{2R_S}\int_0^{\infty}dl\,
\frac{(2l+1)\sqrt{l(l+1)}}{\exp\left(\frac{\beta\sqrt{l(l+1)}}{R_S}
\right)-1}\, .
\ee
The integral is dominated by $l\sim R_S/\beta$; in our case $\beta\sim
1$ (see below) and therefore $R_S/\beta\gg 1$. Then we can approximate
$l(l+1)\simeq l^2$ and we find that
 the sum over modes is the same as in the Rindler case,
with now $A=4\pi R_S^2$.

The value of $\beta$ which enters in the above expressions is the
value of the {\em local} inverse temperature,
since this is the temperature
felt by the membrane. By definition, the membrane
is a coarse graining over  the region
where new physics sets in. As discussed in sect.~3, the
distances where the local Hawking temperature becomes of order one are
$r= R_S+O(\lpl^2/R_S)$. Then, let us consider  a membrane which
represents a coarse graining over the region
\be
r\leq R_S+a\frac{\lpl^2}{R_S}\, ,
\ee
with $a$ a numerical constant to be discussed below.
The corresponding local inverse
temperature is $\beta\simeq 4\pi\sqrt{a}$. Then, for the entropy we
get
\be\label{gamma}
S=\gamma\, \frac{A}{4},\hspace{7mm}
\gamma =\left( \frac{3 c(\Lambda )}{8\pi^2}\right) \frac{1}{\pi a}\, .
\ee
This result is very similar to the result obtained  by
't~Hooft~\cite{tH} and  by Frolov and
Novikov~\cite{FN}, with  approaches which, at first sight, are very
different from ours.
The relation with their result and with the total black hole entropy
is discussed in the next subsection.

\subsection{The total entropy}

While the computation of the entropy of the $\phi$ field is
technically straightforward, its interpretation and
its relation to the total black hole
entropy is not immediate. A number of authors, see
e.g.~\cite{tH,FN,Yor}, have presented  computations in which
some form of mode counting is performed. In all these computations
the resulting entropy
diverges unless one introduces a cutoff near the horizon, like our
constant $a$ introduced in the previous subsection, and the divergence
is $\sim 1/a$. One can fix the cutoff by requiring that it gives just the
standard value for the proportionality constant, $S=A/4$.

{}From the point of view of the
effective Lagrangian approach, however, a different
interpretation seems more natural. In the membrane approach, the
result $S=A/4$ can be obtained from a `zeroth order'
term~\cite{MM2}, and the contribution of the $\phi$ field is more
naturally interpreted as a correction to it.  The reasoning presented in
ref.~\cite{MM2} is as follows. We want to define a path integral over
the quantum field $g_{\mu\nu}$ in the presence of a
black hole (or, more generally, of a horizon), from the point
of view of an external, static observer. In intuitive
terms, we would like to limit ourselves to integration variables which
live outside the horizon. However,
very close to the horizon we have to face
the problem that we are entering the region of super-Planckian
temperatures. To cope with
this difficulty we  divide the space outside the
black hole into the region $r< R_S+a(\lpl^2/R_S)$, and  the region
$\Omega$ defined by $r\geq R_S+a(\lpl^2/R_S)$. The partition function in
the region $\Omega$ is simply
\be
Z_{\rm grav}=\int_{\Omega}{\cal D}g_{\mu\nu}\, e^{iS_{\rm grav}}\, ,
\ee
where $S_{\rm grav}$ is
the standard Einstein-Hilbert action supplemented by the
boundary term on $\partial\Omega$.
However, at leading order in the large $M$ expansion we can
neglect the quantum
fluctuations of the metric $g_{\mu\nu}$ in the region
$\Omega$, and we simply use the classical Schwarzschild metric.
This also implies that, at this order, we do not have to worry about
the measure of integration over $g_{\mu\nu}$, and we write the
Euclidean partition function of the region $\Omega$ as
\be
Z_{\rm grav}\simeq
e^{-S_{\rm grav}}|_{g_{\mu\nu}=g_{\mu\nu}^{\rm cl}}\, .
\ee
On the classical metric $g_{\mu\nu}^{\rm cl}$, the volume term in
$S_{\rm grav}$ is zero and only the boundary term survives.
In the region $r< R_S+a\lpl^2/R_S$ we use instead our effective
action. If we call $Z_{\rm eff}$ the
partition function of the $\phi$ field,
 the total partition function is
\be
Z=Z_{\rm grav} Z_{\rm eff}\, .
\ee
If now we fix the local inverse temperature $\beta$ at the  standard
value, we can compute the free energy $F$ from $Z=e^{-\beta F}$, and
therefore the entropy $S$.
As shown in ref.~\cite{MM2}, the boundary term in the
gravitational action gives just the result $S=A/4$. The computation is
formally the same as the well-known computation of Gibbons and Hawking
in Euclidean quantum gravity~\cite{GH}. The only difference is that we
are evaluating the boundary term on the surface $r=R_S+a\lpl^2/R_S$
while in ref.~\cite{GH} it is computed on a surface at
infinity. However, as shown by York~\cite{Yor2}, if one  computes the
derivatives of the thermodynamic potentials with respect to the local
inverse temperature $\beta$, rather than with respect to the inverse
of the temperature at infinity, the result for the entropy
is independent of the
position of the surface used. This explains why we obtain the
same result as the one in ref.~\cite{GH}.

The contribution to the entropy which comes from
 $Z_{\rm eff}$ is instead what we have computed
in the previous subsection. Therefore we find, for the black hole
entropy $S_{\rm bh}$
\be
S_{\rm bh}=\frac{1}{4}(1+\gamma )\, A\, .
\ee
It remains to estimate the order of magnitude of $a$ and therefore of
$\gamma$. In our approach $a$ is not a cutoff put in by hand,
and to be removed with a renormalization procedure, but it is a number
fixed by physics, and it depends on the scale at which new physics
sets in. If new physics sets in at the Planck scale, the constant $a$
is, parametrically, of order one. Thus $\gamma$ is not parametrically
small compared to one, even if, for typical values of $a$ and
$\Lambda$, it might be numerically small.

If however superstrings are the fundamental theory, new physics sets
in at the string scale. If we denote by $\lambda_s$ the string length,
we start to perform the coarse graining when the local temperature
$T$ reaches a value
$\sim 1/\lambda_s$, or $\beta\sim\lambda_s$. Since $\beta$
is related to $a$ by $\beta \simeq 4\pi\sqrt{a}$, this gives (writing
explicitly also the Planck length $\lpl$),
\be
a=O\left(\frac{\lambda_s^2}{\lpl^2}\right)\, .
\ee
(Note that in heterotic string theory this means $a\sim\alpha_{\rm
GUT}^{-2}$, since $\lpl /\lambda_s =\alpha_{\rm GUT}/4$.)
Thus  $\gamma\sim 1/a$ is a parametrically small
quantity, and we get
\be
S=\frac{A}{4}\, \left( 1+ O(\frac{\lpl^2}{\lambda_s^2})\right)\, .
\ee
The interpretation of the contribution of the $\phi$ field as a
parametrically small correction to the entropy also allow us to see in
a different light the results of refs.~\cite{tH,FN}, where
it is computed the
contribution of external matter fields to the entropy. If matter fields
$\Phi$ are present, we would write for the Euclidean partition
function $Z=Z_{\rm grav} Z_{\rm eff}Z_{\Phi}$,
where $Z_{\Phi}$ is the partition function
 of the matter fields restricted to the region $\Omega$.
Their contribution to
the entropy  has been computed by
 't~Hooft using a brick wall regulator, and the horizon contribution is
$(A/4)(Z/(360\pi a))$, where we have written 't~Hooft regulator $h$ as
$h=a/R_S$, $Z$ is the number of fields,
and we have used the standard value of the temperature.
Note that this is identical, even in the numerical
constant, to the result found by Frolov and Novikov~\cite{FN} with a
different method.
If, instead of choosing $a=1/(360\pi )$ in order to reproduce the
black hole entropy, we rather say that $a$ is fixed by the scale where
new physics sets in, and we identify this scale with $\lambda_s$,
then this contribution is just another
 parametrically small correction to the black hole entropy.

This interpretation allows us to get rid of some inconsistencies
which are inherent in the attempt to identify the contribution of
matter fields with the total black hole entropy. In particular the
horizon contribution to the internal energy
becomes parametrically smaller than $M$, while if one choses
$a=1/(360\pi )$ one gets~\cite{tH} $U=(3/8)M$, which is a sizeable
fraction of the total black hole mass.
The same happens of the specific heat, which with the choice
$a=1/(360\pi )$
 is positive and even larger, in module, than the
black hole value $-8\pi M^2$~\cite{BL}.
Another hint in favor of this interpretation comes from the study of
the brick wall model for a black hole with charge $Q\neq 0$. With a
straightforward repetition of  't~Hooft computation for the case
$Q\neq 0$ we have found for the entropy of near-extremal
black holes $S\sim (A/a)(1-(Q^2/M^2))^{1/2}$. If we want to fix $h=a/R_S$
in such a way as to obtain the result $S=A/4$ we must chose $h\sim
(1-(Q^2/M^2))^{1/2}/R_S$, which is quite inconsistent, since
for near-extremal black hole it is well
within the region of super-Planckian temperatures. If, on the
contrary, we chose $h$ as the scale at which string physics comes in,
we simply obtain a correction to the $A/4$ result which is suppressed
both by the small parameter $(\lpl /\lambda_s)^2$ and by the factor
$(1-(Q^2/M^2))^{1/2}$.

\section{Conclusions}
Following the pioneering work of 't~Hooft~\cite{tH}, a number of
authors have recently considered the possibility that, in spite of the
fact that in classical general relativity the horizon merely
represents a coordinate singularity, at the quantum level it is  the
place where the quantum degrees of freedom of black holes should be
found, at least from the point of view of a static observer.

In this approach,  when we are sufficiently close  to
 the horizon we are  entering a region of very high
temperatures (because of the blue-shift factor in the Hawking
temperature), and unknown physics sets in. Thus, to describe  properly
the region close to the horizon is a very difficult and
essentially non-perturbative problem. On the
other hand, it is a very important problem, which might be the key to
understanding the statistical origin of black hole
entropy and the information loss paradox.

In order to cope with our ignorance of short distance physics
 we have proposed an effective Lagrangian approach.
In this paper we have
investigated the general formalism which follows from the 'Basic
Postulate' given in the Introduction.  We have shown how to
construct the most general effective action, and we have found that
the various terms are organized  as an expansion in powers of $\mpl /M$.
We have discussed a
 semiclassical expansion of the action which is in a sense
complementary to the treatement given in ref.~\cite{MM1}.
In ref.~\cite{MM1} the field theory resulting from the membrane action
was truncated to the radial mode, and the resulting quantum mechanical
problem was discussed exactly, i.e., without performing a
semiclassical expansion. In this paper, instead, we have retained all
modes of the membrane, but we have used a semiclassical approximation.

Probably the main message of this paper is that at this
effective Lagrangian level the region very close to the horizon
can be studied by rather standard field theoretical methods. The
fluctuations of the horizon are described by a single scalar field
living in a 2+1 dimensional curved space, governed by an action which is
completely fixed, apart from a number of phenomenological
constants, like the membrane tension, in which it is summarized all
our ignorance of short distance physics.

With this approach it is possible to compute the black hole entropy,
and we find the result $S=A/4$ plus corrections which depend on the
scale at which new physics sets in.

\vspace{5mm}

Acknowledgments. We thank Enore Guadagnini,
Pietro Menotti and Michele Mintchev
for  very useful discussions.
We thank the referee for pointing out an error in the manuscript.

\vspace{5mm}

{\bf Appendix A. The propagator in Rindler space.}

\noindent
Let us compute the propagator $D(\xi ,\xi ')$
of the field $\phi(\xi )$ in the background given by eq.~(\ref{back}).
In terms of the variable $\tilde{\tau}=z_0\tanh g\tau$,
 eq.~(\ref{back}) becomes simply
$ds^2=-d\tilde{\tau}^2+dx^2+dy^2$.
Since $\tilde{\tau}=\pm z_0$
corresponds to $\tau =\pm\infty$, we impose the boundary conditions
$D(\tilde{\tau} =\pm z_0,\tilde{\tau}';{\bf x-x'})=
D(\tilde{\tau} ,\tilde{\tau}'=\pm z_0;{\bf x-x'})
=0$, where we have used the
notation ${\bf x}=(x,y)$.
Thus we are dealing with a massless propagator with
time boundary conditions, in flat space.
 We compute it following ref.~\cite{CM},
with some simple modifications due to the fact that in our case the
boundaries are in Minkowski, rather then in Euclidean space.
It is convenient to consider the Green's function
$D(\tau ,\tau ';{\bf k})$ obtained by
performing the Fourier transform only with respect to
the spatial coordinates ${\bf x}$. A Green's function with the
$\epsilon$ prescription appropriate for the Feynman propagator, but
which does not obey the boundary conditions, is
\be\label{D0}
D_F^{(0)}(\tilde{\tau},\tilde{\tau}';{\bf k})
=\frac{1}{2z_0}\sum_{n=-\infty}^{\infty}
\frac{e^{-i\omega_n (\tilde{\tau}-\tilde{\tau}')}}
     {\omega_n^2-{\bf k}^2-i\epsilon}\, ,
\hspace{15mm} (\omega_n =\frac{\pi n}{z_0})\, .
\ee
The Green's function which vanishes at the time boundaries is then
\be\label{DF}
D_F(\tilde{\tau},\tilde{\tau}')=
D_F^{(0)}(\tilde{\tau},\tilde{\tau}')-
\frac{D_F^{(0)}(\tilde{\tau},z_0 )D_F^{(0)}(z_0 ,\tilde{\tau}')}
{D_F^{(0)}(z_0,z_0 )}\, .
\ee
Using
$D_F^{(0)}(\tilde{\tau},z_0 )=D_F^{(0)}(\tilde{\tau},-z_0 )$
we see that $D_F(\tilde{\tau},\pm z_0 )=D_F(\pm z_0,\tilde{\tau}')=0$.
The sum in eq.~(\ref{D0}) can be performed explicitly and,
expressing the result in terms of $\tau ,\tau '$, we get
\bees
& & D_F({\tau},{\tau}';{\bf k})=
-\frac{1}{|{\bf k}|\sin (2z_0|{\bf k}|)}\times\nonumber\\
& &\left[\theta (\tau -\tau ')
\sin \left( z_0|{\bf k}|\tanh (g\tau ) -z_0|{\bf k}|\right)
\sin \left( z_0|{\bf k}|\tanh (g\tau ') +z_0|{\bf k}|\right)
\right. \nonumber\\
& & \left. +(\tau \leftrightarrow\tau ')\right]\, .\label{a1}
\ees
Here $|{\bf k}|$ actually is $|{\bf k}|+i\epsilon$.
This result can also
be directly obtained from eq.~(B.2) of ref.~\cite{CM},
where the boundary conditions are in Euclidean space,
with the formal replacement ${\bf k}\ra i{\bf k}$.

The field theory obtained by expanding
around a generic, non-planar, solution of the equations of motion is
qualitatively similar to that obtained expanding about
a planar solution; in fact,
it is easy to check from the equations of motion
that, for an arbitrary solution, any 'bump' is smoothed out
exponentially with $\tau$.  Thus,  asymptotically
$z_{\rm cl}\ra z_0/\cosh (g\tau )$,
independently of the initial conditions,
and therefore the background metric is still given,
asymptotically,  by eq.~(\ref{back}).

\vspace{5mm}

\vspace{5mm}

{\bf Appendix B. The black hole in 2+1 dimensions}

\noindent
It might be interesting to examine our formalism  also for the
2+1 dimensional black hole solution discovered by
Ba\~nados, Teitelboim and Zanelli (BTZ)~\cite{BTZ}.
In spite of the many differences between three-dimensional and
four-dimensional gravity, the BTZ black hole has remarkable similarities
with its four-dimensional analog, and a number of investigations of
its geometrical and thermodynamical properties have appeared
recently, see e.g.~\cite{BTZ,BTHZ,ER,Car} and references therein.
One considers the theory defined by
the action
\begin{equation}\label{action}
I_{\rm grav}=\frac{1}{16\pi G}
\int d^3x\, \sqrt{-g}\left[ {\cal R}+2l^{-2}\right] +B
\end{equation}
where $l$ is related to the cosmological constant $\Lambda$ by
$\Lambda =-l^{-2}<0$
and $B$ is the boundary term. Writing the metric
in the ADM form (and setting to zero the shift functions $N^i$),
$ds^2=N^2dt^2- g_{ij}dx^idx^j$,
the boundary term is given by
\begin{equation}
B=-\frac{1}{8\pi G}\int d^3x\,\partial_i\left[
g^{ij}\partial_jN\, \sqrt{^{(2)}g}\,\right]\, ,
\end{equation}
where $^{(2)}g={\rm det}\, g_{ij}$. Hereafter, following~\cite{BTZ},
we will use units $G=1/8$. As before,
the embedding spacetime curvature is denoted by ${\cal R}$ while
 $R$ will be used for the world-sheet curvature.
The BTZ black hole, limiting ourselves to zero angular momentum, is
given by
\begin{equation}
ds^2=-\alpha dt^2 +\alpha^{-1} dr^2 +r^2d\theta^2\, ,
\hspace{7mm}
\alpha = \frac{r^2-r_+^2}{l^2}\, ,
\end{equation}
where $r_+=l\sqrt{M}$. In this case the
collective coordinates $\zeta^{\mu}$ depend on two variables
$\xi^i=(\tau ,\sigma )$. Instead of a membrane, the horizon is
now described by a string, with induced metric $h_{ij}$ and extrinsic
curvature $K_{ij}$.  Embedding
space indices now take values $\mu =0,1,2$ and world-sheet indices take
values $i=0,1$.
The effective action can be written using the same arguments as in
sect.~2. However, now
we have some simplifications; at order $n=2$ the
term proportional to the world-sheet curvature does not
contributes since $(-h)^{1/2}R$ is a total derivative. At
this order, therefore, the most general effective action is
\begin{equation}
S=-{\cal T}\int d^2\xi \,\sqrt{-h}\, \left[ 1+C_0K+C_I K^2
+C_{II}K_i^jK^i_j\right]\, .
\end{equation}
On the BTZ metric the terms $K^2$ and $K_i^jK^i_j$ are not independent
and we can restrict ourselves to one of them. When $C_0=0$
this is the action for the rigid string  discussed by
Polyakov~\cite{Pol}.
The equation of motion at order $n=0$ is again $K=0$.
Defining $y=\alpha^{1/2}/g$
and $g=r_+/l^2=\sqrt{M}/l$, the equation of motion $K=0$
for a circular string reads
\begin{equation}\label{11}
y\ddot{y}-2\dot{y}^2+g^2y^2 +(\frac{2g^2}{l^2})y^4
-2\frac{y^2\dot{y}^2}{l^2+y^2}=0\, .
\end{equation}
This equation can be integrated exactly.
With the initial conditions $y(0)=al,\dot{y}(0)=0$, where $a$ is a
dimensionless parameter, the exact solution is
\begin{equation}\label{solut}
g\tau = \frac{1}{a}\,\sqrt{\frac{1+a^2}{1+2 a^2}}
\, \Pi \left(\psi ,\frac{1+a^2}{a^2},
\sqrt{\frac{1+a^2}{1+2 a^2}}\,\right)\, .
\end{equation}
Here $\Pi (\psi ,\alpha^2,k)$ is the elliptic integral of the third
kind, and $\sin^2\psi = (a^2-(y/l)^2)/(a^2+1)$.
However, as in the 3+1 dimensional case, it is more convenient to
consider the Rindler limit rather than working with the exact solution.
Let us examine the
Rindler limit for the BTZ black hole. In terms of $y=\alpha^{1/2}/g$,
and of $x$ defined by
$x=r_+\theta$, the BTZ metric can be written as
\begin{equation}
ds^2=-g^2y^2dt^2+(1+\frac{y^2}{l^2})^{-1}dy^2+(1+\frac{y^2}{l^2}) dx^2
\, .
\end{equation}
If we are sufficiently close to the horizon, $y\ll l$, we get a $2+1$
dimensional Rindler metric,
$ds^2=-g^2y^2 dt^2 + dx^2 + dy^2$. It is interesting to observe that,
since $l=\sqrt{M}/g$, the large $l$ limit can be obtained by taking
the black hole mass $M$ large, with $g$ held fixed and arbitrary,
while in  $3+1$ dimensions
the Rindler limit necessarily implies $g\rightarrow 0$.
In the $2+1$ Rindler metric the  equation of motion of
the string reads
$y\ddot{y}-2 \dot{y}^2 +g^2y^2=0$,
which is the same as in  the
$3+1$ dimensional case.

The semiclassical expansion can be performed in analogy with the 3+1
dimensional case, and the
effective action, at quadratic order, is formally identical to
eq.~(\ref{lin}), with $d^3\xi\ra d^2\xi$. The computation of the
entropy of the $\phi$ field proceeds
along the lines of sect.~4 and gives a result proportional to the
perimeter of the horizon. The zeroth order term, instead, has already
been computed in ref.~\cite{MM3}:
we define the local inverse temperature
$\beta =\beta_{\infty}\alpha^{1/2}$ and we compute
the boundary term on a
spherical surface with a generic radius $r$. Considering $M$ as a
function of $\beta$ and $r$ defined implicitly by
$\beta =\beta_{\infty}\alpha^{1/2}$, we find $B=B(\beta ,r)$, and the
entropy is given by
\begin{equation}
S=\beta\left(\frac{\partial B}{\partial\beta}\right)_r-B\, .
\end{equation}
A simple computation gives
\begin{equation}
B=-2\beta_{\infty}\frac{r^2}{l^2}=-4\pi r
\left[ 1+(\frac{\beta}{2\pi l})^2\right]^{1/2}\, ,
\end{equation}
and therefore
\begin{equation}
S=4\pi r_+ = 2\times {\rm perimeter}\, ,
\end{equation}
in agreement with ref.~\cite{BTZ}.

\end{document}